\documentclass[12pt]{iopart}

\usepackage[colorlinks=true,urlcolor=blue,citecolor=blue,linkcolor=blue]{hyperref}

\usepackage{graphicx}
\usepackage{caption}
\usepackage{color}
\usepackage{bm}
\usepackage{dsfont}

\makeatletter
\renewcommand\@appendixstar{\@@par
 \ifnumbysec 
 \@addtoreset{table}{section}
 \@addtoreset{figure}{section}\fi
 \setcounter{section}{0}
 \setcounter{subsection}{0}
 \setcounter{subsubsection}{0}
 \setcounter{equation}{0}
 \setcounter{figure}{0}
 \setcounter{table}{0}
 \def\thesection{\Alph{section}} 
 \def\theequation{\ifnumbysec
 \Alph{section}.\arabic{equation}\else
 \Alph{section}\arabic{equation}\fi}
 \def\thetable{\ifnumbysec
 \Alph{section}\arabic{table}\else
 A\arabic{table}\fi}
 \def\thefigure{\ifnumbysec
 \Alph{section}\arabic{figure}\else
 A\arabic{figure}\fi}}
\makeatother

\begin{document}

\title{Quantum optimization using variational algorithms on near-term quantum devices}

\author{Nikolaj~Moll$^1$, Panagiotis~Barkoutsos$^1$, Lev~S.~Bishop$^2$, Jerry~M.~Chow$^2$, Andrew Cross$^2$, Daniel~J.~Egger$^1$, Stefan~Filipp$^1$, Andreas~Fuhrer$^1$, Jay~M.~Gambetta$^2$, Marc~Ganzhorn$^1$, Abhinav~Kandala$^2$, Antonio~Mezzacapo$^2$, Peter~M\"uller$^1$, Walter~Riess$^1$, Gian~Salis$^1$, John~Smolin$^2$, Ivano~Tavernelli$^1$, and Kristan~Temme$^2$}
\address{$^1$ IBM Research -- Zurich, S\"aumerstrasse 4, 8803 R\"uschlikon, Switzerland}
\address{$^2$ IBM T.J. Watson Research Center, Yorktown Heights, NY 10598, USA}

\date{\today}

\begin{abstract} 
Universal fault-tolerant quantum computers will require error-free execution of long sequences of quantum gate operations, which is expected to involve millions of physical qubits. Before the full power of such machines will be available, near-term quantum devices will provide several hundred qubits and limited error correction. Still, there is a realistic prospect to run useful algorithms within the limited circuit depth of such devices. Particularly promising are optimization algorithms that follow a hybrid approach: the aim is to steer a highly entangled state on a quantum system to a target state that minimizes a cost function via variation of some gate parameters. This variational approach can be used both for classical optimization problems as well as for problems in quantum chemistry. The challenge is to converge to the target state given the limited coherence time and connectivity of the qubits. In this context, the \emph{quantum volume} as a metric to compare the power of near-term quantum devices is discussed.  

With focus on chemistry applications, a general description of variational algorithms is provided and the mapping from fermions to qubits is explained. Coupled-cluster and heuristic trial wave-functions are considered for efficiently finding molecular ground states. Furthermore, simple error-mitigation schemes are introduced that could improve the accuracy of determining ground-state energies. Advancing these techniques may lead to near-term demonstrations of useful quantum computation with systems containing several hundred qubits.
\end{abstract}

\pacs{quantum computation, quantum chemistry, quantum algorithms}

\maketitle

\tableofcontents

\title[]{} 

\section{Introduction}  

Recent advances in the field of quantum computing have boosted the hope that one day complex problems can be solved efficiently on quantum computers. The ultimate goal is a universal fault-tolerant quantum computer that runs arbitrary algorithms much faster than on a classical computer. However, millions of physical qubits and high-fidelity gate operations are required to implement a universal fault-tolerant quantum computer, a system that currently cannot be built. Yet, quantum devices with a couple of hundred physical qubits with limited or no error correction are likely to become available in the near future. With it comes the question how to exploit these devices for useful calculations. In this paper, we discuss how the variational quantum eigensolver can be run on near-term quantum devices to tackle optimization problems that are exponentially hard on classical computers.

We differentiate between two types of optimization problems. The first kind are quantum optimization problems, such as finding the ground state of a complex molecule or the simulation of its dynamics. In this case, optimization typically involves minimization of the total energy as described by the energy expectation value of a non-trivial Hamiltonian as a function of some molecular parameters, such as interatomic distances. The second kind are classical optimization problems which can usually be mapped onto a relatively simple Ising-type Hamiltonian. In both cases, exponential scaling of the required computational resources with the problem size can make the problems hard to solve or even in-tractable on classical computers. 

Generally, optimization problems are solved by finding the extremum of an objective function, such as cost, energy, profit or error. As the cost function typically depends on a large set of parameters, finding a solution involves searching a high-dimensional parameter space, which quickly makes a brute-force approach unfeasible. A quantum computer operates on Hilbert space, which grows exponentially as $2^N$ with the number of qubits $N$. The idea is to use this vast state space with the help of quantum entanglement, and thus boost the efficiency in finding the right solution, ideally with exponential speed-up~\cite{lloyd_universal_1996, abrams_simulation_1997, abrams_quantum_1999, aspuru-guzik_simulated_2005, harrow_quantum_2009}. A more careful analysis shows, however, that the speed-up for classical optimization problems is in many cases rather modest~\cite{denchev_whatis_2016, albash_evidence_2017, smolin_classical_2014}. In contrast, one can benefit from quantum speed-up in problems that are directly related to the quantum-mechanical description of nature itself. A prominent example is finding the many-electron wavefunction of a molecular system. Classical computers fail to solve such problems exactly for more than a few tens of electrons because of the exponential increase of Hilbert space with the number of electrons. The large state space of a quantum computer can be used to simulate a chemical system and calculate its properties, including correlations and reaction rates, once the challenge of efficiently mapping the fermionic problem to the available qubit hardware is overcome. 

In fact, on a quantum device the natural way is to solve the chemical system in second quantization~\cite{abrams_quantum_1999, aspuru-guzik_simulated_2005, buluta_quantum_2009, lanyon_towards_2010, brown_using_2010, temme_quantum_2011, kassal_simulating_2011, whitfield_simulation_2011, aspuru-guzik_photonic_2012, whitfield_computational_2012, yung_quantumquantum_2012, jones_faster_2012, toloui_quantum_2013, wecker_gate-count_2014, georgescu_quantum_2014, hastings_improving_2015, poulin_trotter_2015, mueck_quantum_2015, garcia-alvarez_quantum_2015, whitfield_unified_2015, reiher_elucidating_2017, babbush_exponentially_2016, whitfield_local_2016, wendin_quantum_2016, popkin_quest_2016, shen_quantum_2017, colless_robust_2017} formulated in terms of fermionic annihilation and creation operators. Because of the different statistics there is no direct one-to-one mapping: each fermion operator must be represented by a string of qubit operators, which induces long-range qubit-qubit correlations in the system and places demanding requirements on the connectivity and the number of gates (see Section~\ref{sec:mappingqubits}). To compute the quantum evolution of chemical systems on a digital quantum computer, decomposition into discrete time steps is required and accordingly long gate sequences \cite{abrams_quantum_1999, whitfield_simulation_2011, omalley_scalable_2016}.

\begin{figure}[tb]
\begin{center}
\includegraphics[width=152mm]{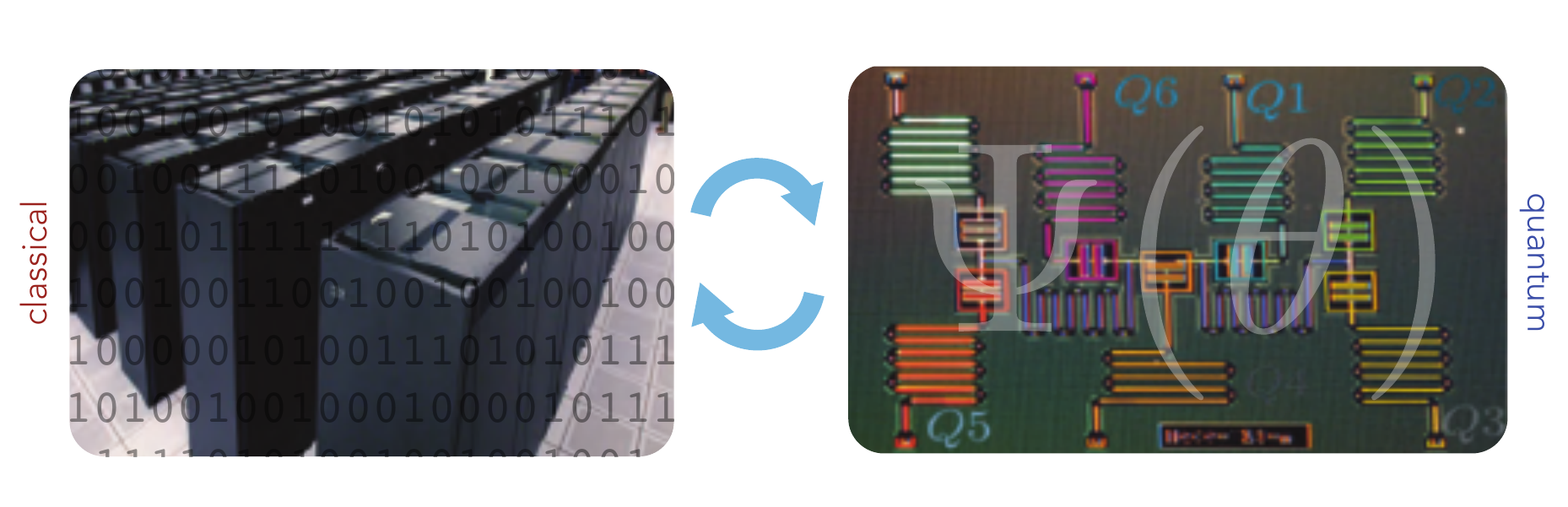}
\end{center}
\caption{Schematic of a hybrid quantum classical computing architecture.}
\label{fig:hybrid}
\end{figure}

On current quantum devices, gate errors and decoherence restrict the number of sequential gate operations that can be performed while keeping a meaningful, coherent quantum state. Moreover, connectivity between qubits is limited by the physical routing of the wires on a qubit chip. This is why a new class of hybrid classical quantum algorithms, called the variational quantum eigensolver (VQE)~\cite{li_solving_2011, tempel_quantum_2012, peruzzo_variational_2014, omalley_scalable_2016, wecker_progress_2015, li_efficient_2017, mcclean_theory_2016, colless_robust_2017,farhi_quantum_2014, farhi_quantum_2014-1, farhi_quantum_2016}, holds a lot of prospects for near-term quantum-computing systems (see Fig.~\ref{fig:hybrid}).  
These algorithms work with short-depth circuits and will result in approximate results when the number of qubits, their coherence and the connectivity is large enough. These requirements on the quantum system can be quantified by the \emph{quantum volume}~\cite{bishop_quantum_2017}, a hardware-independent figure or merit for the power of a quantum computer. 

The VQE can be used both for classical optimization problems as well as for fermionic Hamiltonians describing, e.~g., quantum chemistry. In quantum chemistry the variational quantum eigensolver is used to calculate ground states~\cite{li_solving_2011, tempel_quantum_2012, peruzzo_variational_2014, omalley_scalable_2016, wecker_progress_2015, li_efficient_2017, colless_robust_2017} of chemical systems. The high-dimensional trial wavefunctions, which are costly to represent on a classical computer, are generated on the quantum computer using parametrized single-qubit and entangling gates. The optimization of the gate parameters is performed on a classical computer by summing expectation values of the qubit operators measured on the quantum device and thereby calculating the total energy as a cost function. This can in principle lead to very short-depth circuits which ideally run in a time that is shorter than the coherence time of the quantum computer. The same variational quantum eigensolver can be applied to other physical systems in condensed matter such as the Fermi-Hubbard model~\cite{abrams_simulation_1997, verstraete_mapping_2005, temme_quantum_2011, yung_quantumquantum_2012, barends_digital_2015, reiner_emulating_2016, havlicek_operator_2017} and spin systems~\cite{heras_digital_2014, you_quantum_2010, biamonte_nonperturbative_2008, zintchenko_local_2015}.

Hybrid algorithms are, however, not resilient against decoherence and gate errors, which may lead to inaccurate estimates of the expectation values. Currently available error-correction schemes, such as those based on surface codes \cite{fowler_surface_2012}, require a significant number of qubits, rendering quantum simulations of practical systems challenging in the near future.  Still, novel schemes that do not require ancillas or code qubits can help mitigate induced errors, enabling longer and bigger quantum computations. Such error mitigation schemes \cite{schwenk_reconstructing_2017, temme_error_2016} need to be developed further and tested to improve accuracy without the full overhead of error-correction codes for universal quantum computing.  

The paper is structured as follows: The quantum volume is discussed in Section~\ref{sec:qv} before we explain the variational quantum eigensolver in Section~\ref{sec:vir} and its application to quantum chemistry problems in Section~\ref{sec:chemistry}. After a brief discussion of the prospects of solving classical optimization problems with near-term quantum devices in Section~\ref{sec:qaoa}, we elaborate on the choice of suitable optimizers for the classical feedback in the VQE in Section~\ref{sec:opt} and discuss the prospects of fighting back decoherence in near-term quantum devices without full error correction in Section~\ref{sec:error}. Finally, we conclude in Section~\ref{sec:conclusion}.

\section{Quantum volume, a metric for near-term quantum devices}
\label{sec:qv}

For current quantum processors, various architectures and physical qubit realizations are being considered. While quantum systems based on superconducting qubits~\cite{corcoles_demonstration_2015, takita_demonstration_2016, riste_detecting_2015, omalley_scalable_2016, barends_superconducting_2014, ofek_extending_2016, qx_ibm_quantum_2016} at the moment seem to be leading the way, ion-trap-based systems~\cite{debnath_demonstration_2016, monz_realization_2016} are close competitors. Furthermore, semiconductor-based spin qubits~\cite{veldhorst_two-qubit_2015, zajac_quantum_2017, nichol_high-fidelity_2017} and other quantum architectures~\cite{mourik_signatures_2012, sarma_majorana_2015, obrien_optical_2007} may still become important in the future. Given the different hardware implementations it is often difficult to benchmark the usefulness or power of quantum systems, which is why a hardware-independent measure is required. To define a suitable metric, we first note that a quantum computer's performance depends on five main hardware parameters: 

\begin{enumerate}
\item Number of physical qubits $N$
\item Connectivity between qubits
\item Number of gates that can be applied before errors or decoherence mask the result
\item Available hardware gate set
\item Number of operations that can be run in parallel
\end{enumerate}
With the goal to quantify a quantum computer's power with a single parameter, we would like to consider a metric based on the question `can this device run a given algorithm?'. For any given instance of a quantum algorithm, there is a lower bound on the number of qubits $N$ required to run the algorithm, as well as the necessary number of steps (or circuit depth) $d$. We therefore define a {\it quantum volume} $V_{\rm Q}$~\cite{bishop_quantum_2017} that takes into account both the number of qubits $N$ and the allowable depth $d$ of quantum circuits that can be run on a near-term quantum device. In the simplest case, we could just choose the quantum volume to be $d\cdot N$; however, this has some undesirable properties in that it can be gamed in various ways. For example, in many cases the smallest error rates and therefore the largest circuit depth will result from very few qubits, even $N = 2$, as in this case there will be less connectivity and parallelization overhead and fewer issues with crosstalk between qubits. However, clearly $N = 2$ is a completely uninteresting limit. Also the other extreme, where a device has many qubits but little coherence, i.e.\ $d \approx 1$, is not interesting because such a system cannot use entanglement as a resource and calculations become effectively classical. 

We therefore conceptually define the quantum volume as
\begin{equation}
\label{eqn:qv1}
\tilde V_{\rm Q} =  \min\left[N, d(N)\right]^2 \,.
\end{equation}
Here, the number of qubits $N$ is an easily accessible hardware parameter; however, the achievable circuit depth $d(N)$ needs further specification in terms of the hardware parameters given in the list above. 

\begin{figure}[tb]
\begin{center}
\includegraphics[width=140mm]{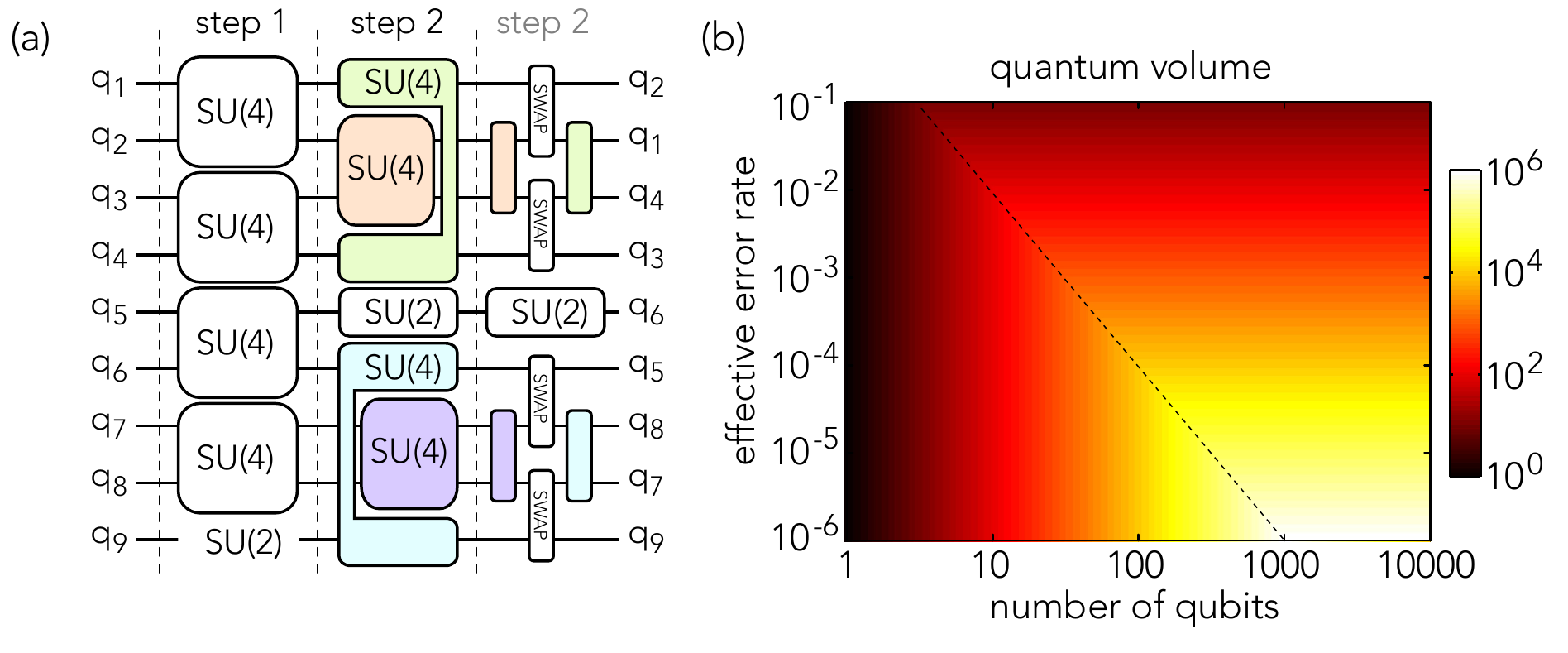}
\end{center}
\caption{(a) Example quantum circuit with two circuit steps. Step 2 requires different connectivity and would lead to an increased gate count on quantum hardware with only nearest neighbor interactions. This is illustrated to the right of step 2.  (b) Quantum volume as a function of the effective error rate $\epsilon_{\rm eff}$ and the physical number of qubits $N$. For simplicity, we assume that   for a given $\epsilon_{\rm eff}$, $V_{\rm Q}$ stays constant for $d < N$ . } 
\label{fig:fig2}
\end{figure}

We start by considering one step of a quantum algorithm (a depth-one circuit) on a number of $N$ qubits. Such a step is expressed as a unitary operator that can be written as a tensor product of randomly chosen arbitrary two-qubit gates on disjoint pairs of qubits (see step 1 in Fig.~\ref{fig:fig2}(a)). Here, we allow any unitary two-qubit operation in the SU(4) group, which may consist of a combination of one- and two-qubit gates on the actual hardware. Then an effective error rate $\epsilon_{\rm eff}$ is defined as the error rate per two-qubit gate averaged over many realizations of such depth-one circuits.  Therefore, $\epsilon_{\rm eff}$ depends on the gate overhead required when all-to-all connectivity, full parallelism and a suitable gate set is not available. Thereby, it also encapsulates both the errors of single- and two-qubit gates. If the hardware supports all possible two-qubit gates directly (requiring an all-to-all connectivity) with identical error rate $\epsilon$, and in addition allows unlimited gate parallelism, then $\epsilon_{\rm eff} = \epsilon$. If the connectivity is limited, then it will be necessary to insert additional SWAP gates to permute the qubits in order to implement the random two-qubit gates, leading to an increase of $\epsilon_{\rm eff} > \epsilon$. A planar nearest-neighbor qubit coupling would lead to an effective error rate of $\epsilon_{\rm eff} \propto \sqrt{N} \epsilon$, and a linear chain of qubits would yield an effective error rate of $\epsilon_{\rm eff} \propto N \epsilon$. On the other hand a hardware which supports more complex gates such as the Tofoli gate directly or the use of a compiler which efficiently compresses the gates of a test circuit could also lead to a situation with $\epsilon_{\rm eff} < \epsilon$. Other special features and limitations of the hardware must be dealt with in a similar manner.

The error rate of a single circuit step scales with the number of simultaneous two-qubit gates $\epsilon_{\rm 1step}\propto N \epsilon_{\rm eff}$. In other words, we can estimate the circuit depth in which, on average, a single error occurs as $d \simeq 1/(N \epsilon_{\rm eff})$, linking the effective error $\epsilon_{\rm eff}$ to the previous definition of the quantum volume using the circuit depth. As an example, if an effective error rate $\epsilon_{\rm eff} = 10^{-4}$ is experimentally achievable, depth $d = 10$ algorithms could be run on a 1000-qubit device, and  $d=100$ algorithms on a 100-qubit device. 

However, the effective error rate $\epsilon_{\rm eff}$ will depend not only on the gate error rates and the connectivity but, more generally, on the complexity of the quantum system which grows with the number of qubits, for example, because of crosstalk. The effective error rate $\epsilon_{\rm eff}(N)$ will therefore likely be a function of $N$ even if full connectivity is available. Moreover, $\epsilon_{\rm eff}$ also depends on the sophistication of the scheduling algorithm responsible for mapping the quantum algorithm considered to the hardware. Both hardware and software improvements will thus impact the effective error rate $\epsilon_{\rm eff}(N)$. 

Finally, we note that with this definition the allowable circuit depth $d \simeq 1/(N \epsilon_{\rm eff})$ decreases with $N$ at constant effective error $\epsilon_{\rm eff}$, which means that a system's quantum volume decreases if more qubits with the same fidelity are made available on the hardware. However, a given algorithm does not necessarily need all $N$ available qubits. It could even be beneficial for an algorithm that requires $n<N$ qubits to run on an $N$-qubit machine when selecting a subset of qubits with good connectivity is selected. We therefore further refine the definition of the quantum volume in Eq.~(\ref{eqn:qv1}):
\begin{equation}
V_{\rm Q} =  \max\limits_{n < N}\left(\min\left[n,\frac{1}{n\epsilon_{\rm eff}(n)}\right]^2\right) \, ,
\end{equation}
where the maximum is taken over an arbitrary choice of $n$ qubits to maximize the quantum volume that can be obtained with such a subset. To illustrate this, we plot an example quantum circuit with two circuit steps and the functional dependence of the quantum volume on the number of qubits and an effective two-qubit error rate in Fig.~\ref{fig:fig2}. 
The dashed line denotes the tipping point where $d = 1/(N \epsilon_{\rm eff}) = N$. From any point on this line, a significant increase in $V_{\rm Q}$ requires improvements in both $\epsilon_{\rm eff}$ and $N$. We also see that the usefulness of current quantum devices is likely limited by the typical effective error rates, which are $\epsilon_{\rm eff} > 10^{-3}$. To improve $\epsilon_{\rm eff}$ we will have to start encoding quantum states in logical qubits with an overhead in the number of physical qubits. This will eventually lead to fault tolerant quantum computing.
 
The quantum volume is therefore an architecture-neutral metric that characterizes the capability of a chosen quantum computing architecture to run useful quantum circuits. It enables the comparison of hardware with widely different performance characteristics and quantifies the complexity of algorithms that can be run on such a system. An important conclusion that we can draw for the usefulness of near-term quantum devices is that when increasing the number of qubits the power of the quantum device will increase only if the effective error rate is improved at the same time.

\section{Exploring Hilbert space with the variational quantum eigensolver}
\label{sec:vir}

To exploit near-term quantum devices, applications and algorithms have to be tailored to current quantum hardware with only tens or hundreds of qubits and without full quantum error correction. One main constraint is the limited quantum volume that restricts the depth of meaningful quantum circuits. Still, a small-scale quantum computer with hundred qubits can process quantum states that cannot even be stored in any classical memory. A natural way to make use of this quantum advantage is via a hybrid quantum-classical architecture: A quantum co-processor prepares multi-qubit quantum states $|\Psi(\bm{\theta})\rangle$ parametrized by control parameters $\bm{\theta}$. The subsequent measurement of a cost function $E_q(\bm{\theta}) = \langle\Psi(\bm{\theta}) | H_q | \Psi(\bm{\theta})\rangle$,  typically the energy of a problem Hamiltonian $H_q$, serves a classical computer to find new values $\bm{\theta}$ in order to minimize $E_q(\bm{\theta})$ and find the ground-state energy
\begin{equation}
\label{eq:Hq}
E_q^{\rm{min}} = \min_{\bm{\theta}} \left(\langle\Psi(\bm{\theta}) | H_q | \Psi(\bm{\theta})\rangle\right) \, .
\end{equation}
This variational quantum eigensolver approach to Hamiltonian-problem solving has been recently applied in different contexts \cite{Barrett2013, peruzzo_variational_2014, omalley_scalable_2016, mcclean_theory_2016, eichler_exploring_2015, kandala_hardware-efficient_2017}. In fact, the Hamiltonian $H_q$ can take many forms, the only requirement being that it can be mapped to a system of interacting qubits with a non-exponentially increasing number of terms. Here we distinguish two relevant cases: Hamiltonians that describe fermionic condensed-matter or molecular system (Section \ref{sec:chemistry}) and Hamiltonians that describe a classical optimization problem (Section \ref{sec:qaoa}).

\subsection{Variational quantum eigensolver method}

\begin{figure}[tb]
\begin{center}
\includegraphics[width=152mm]{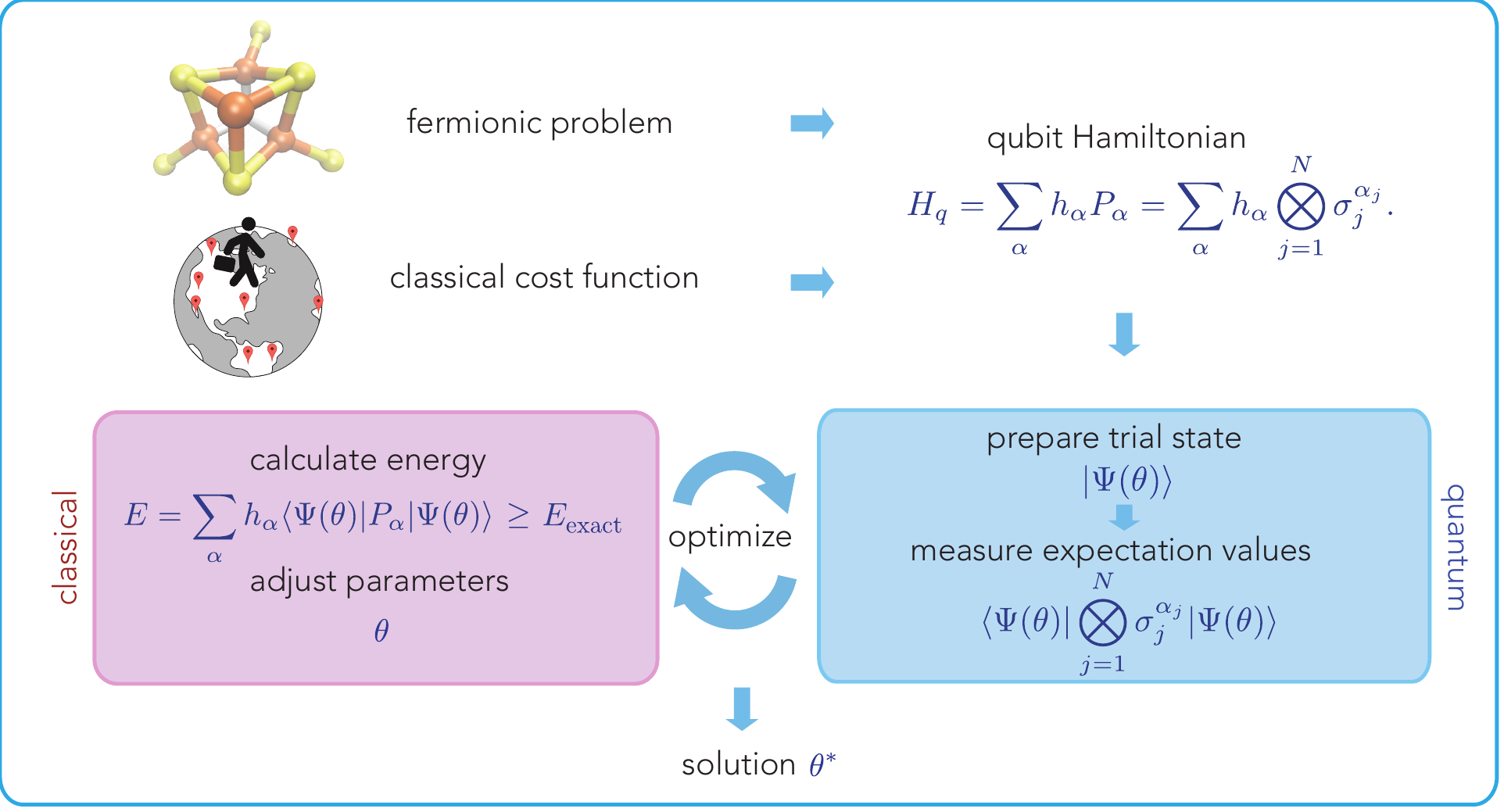}
\end{center}
\caption{Variational quantum eigensolver method. The trial states, which depend on a few classical parameters $\bm{\theta}$, are created on the quantum device and used for measuring the expectation values needed. These are combined on a classical computer to calculate the energy $E_q(\bm{\theta})$, i.e.~the cost function, and find new parameters $\bm{\theta}$ to minimize it. The new $\bm{\theta}$ parameters are then fed back into the algorithm. The parameters $\bm\theta^*$ of the solution are obtained when the minimal energy is reached.}
\label{fig:algorithm}
\end{figure}

In detail, the variational quantum eigensolver method consists of four main steps as shown in Figure \ref{fig:algorithm}.
First, on the quantum processor a tentative variational eigenstate, a trial state, $| \Psi(\bm{\theta}) \rangle$ is generated by a sequence of gates parameterized by a set of control parameters $\bm{\theta}$. In the ideal case, this trial state depends on a small number of classical parameters $\bm{\theta}$, whereas the set of gates is chosen to efficiently explore Hilbert space. In particular, the class of states forming the solution to the minimization problem in Eq.~(\ref{eq:Hq}) has to lie within the set of possible trial states. Suitable gate sets which provide a good approximation to the wanted target state, which minimizes the cost function, have been found for both classical optimization problems \cite{farhi_quantum_2014} (Section \ref{sec:qaoa}) and quantum chemistry problems (Section \ref{sec:chemistry}). Aside from these considerations, it is also essential that hardware constraints be taken into account. As not all gates are directly realizable in hardware, decomposing them into those available in the quantum hardware adds extra overhead in circuit depth. An alternative is, therefore, to use a heuristic approach based on gates that are readily available in hardware \cite{kandala_hardware-efficient_2017} as discussed below. 

Second, once the trial state has been prepared and the expectation value of the problem Hamiltonian $H_q$ is determined. The problem Hamiltonian can be decomposed into Pauli strings $P_\alpha = \sigma_1^{\alpha_1}\otimes \sigma_2^{\alpha_2}\otimes \ldots \sigma_N^{\alpha_N}$ with single-qubit Pauli operators $\sigma_i^j \in \{\mathds{1},\sigma_i^x,\sigma_i^y,\sigma_i^z\}$ and the identity operator $\mathds{1}$, 
\begin{equation}
H_q= \sum_\alpha h_\alpha P_\alpha.
\label{eq:Hqpauli}
\end{equation}
$N$ denotes the number of qubits. To determine the expectation value of each Pauli operator in $P_\alpha$, each single qubit's population is measured repeatedly for a given number of experiments with identical trial state preparation $|\Psi(\bm{\theta})\rangle$. This corresponds to measuring $\sigma_j^z$ for each qubit; other Pauli operators can be determined by applying a pre-rotation on the qubit before the measurement that effectively rotates the measurement axis. To determine the expectation value of the Pauli strings, the measurement outcomes are multiplied for each run of the experiment and then averaged. 

In a third step, the cost function $E_q(\bm{\theta}) = \langle \Psi(\bm{\theta})| H_q | \Psi(\bm{\theta})\rangle  = \sum\limits_\alpha h_\alpha \langle \Psi(\bm{\theta})| P_\alpha | \Psi(\bm{\theta})\rangle $  is calculated by summing up the expectation values of $P_\alpha$ with corresponding coefficients $h_\alpha$. 

Finally, the value of $E_q(\bm{\theta})$ is minimized as a function of the parameters $\bm{\theta}$. A classical optimization algorithm processes $E_q(\bm{\theta})$ and provides new parameters  $\bm{\theta}$. For each parameter set, a new set of gates for trial state preparation has to be loaded onto the quantum processor. As this requires rather time-consuming re-programming of the quantum hardware, it is important that only a minimal number of queries should be made to the quantum processor. Moreover, the calculated expectation values will be noisy because of the limited sampling statistics of the qubit state. Therefore, classical robust optimizers have to used that can handle the noise on the measured expectation values and scale favorably with the number of parameters as described in Section~\ref{sec:opt}. The procedure ends when the minimum of $E_q(\bm{\theta})$ in Eq.~(\ref{eq:Hq}) is reached within a given accuracy and the optimal parameters $\bm\theta^*$ are found.

\section{Quantum chemistry with qubits}
\label{sec:chemistry}

To demonstrate the potential of a quantum processor with limited quantum volume, one needs to consider quantum algorithms that provide a large scaling advantage compared with their classical counterparts. The solution of the electronic structure problem in quantum chemistry belongs to this class: Because of the exponential scaling of the problem, it is impossible to find an exact solution to the Schr\"odinger equation of systems with more than a few tens of electrons on a classical computer. Several approximations have been introduced to access the properties of large-scale systems  with more than 1000 electrons on high-performance computers. The aim is to reach the required accuracy for chemical energies ($\sim 50~\mathrm{meV}$). One approach is to approximate the many-electron Hamiltonian itself using, for example, density-functional theory~\cite{kohn_self-consistent_1965}. There, the original system of interacting electrons is replaced by a fictitious one of non-interacting electrons moving in a modified external potential that allows, at least \textit{in principle}, the original exact solution to be recovered. 

An alternative approach starts from the exact Hamiltonian and attempts to find suitable approximations for the system wavefunction in the many-electron Hilbert space. This calculation can, in principle, be performed either within the first or the second quantization formalism. In first quantization, all spatial integrals have to evaluated on the quantum computer. For this reason, approaches based on second quantization are more suited for first-generation quantum devices. In this case, all spatial integrals are evaluated beforehand on a classical computer, whereas the sampling of the Hilbert space is performed in the orbital configuration space spanned by molecular Slater determinants. This approach maps naturally to the variational method described above (Section \ref{sec:vir}). It starts from the one-electron basis states that are obtained by solving the Hartree-Fock equation. These Hartree-Fock orbitals are then used to construct an anti-symmetrized product wavefunction, the Slater determinant, which is used as a starting point for a perturbative expansion. In this expansion a controlled series of {\it excited} configurations is added until a sufficiently accurate approximation of the ground state is found. 

\subsection{Mapping fermions to qubits}
\label{sec:mappingqubits}

The electronic Hamiltonian in second quantization is given by
\begin{equation}
 H_{F} = \sum_{ij} t_{ij} a^{\dagger} _i a_j+ \sum_{ijkl} u_{ijkl} \, \, a^{\dagger}_i a^{\dagger}_k a_l a_j \, ,
\label{eq:Second_quant_Ham}
\end{equation}
where the operators $a_i^{\dagger}$ and $a_i$ create and annihilate electrons in the $i$-th orbital. The parameters $t_{ij}$ and $u_{ijkl}$ describe the one- and two-electron interactions and can be efficiently computed classically as the overlap integrals of the orbitals in the basis set~\cite{jorgensen_second_1981}. The two-electron term scales at most with the number of orbitals to the fourth power~\cite{aspuru-guzik_simulated_2005, babbush_low_2017} and does not grow exponentially, which would prohibit efficient computation even on a quantum computer.

Because $a_i$ and $a_i^{\dagger}$, unlike the Pauli spin operators, follow fermionic commutation rules $\{a_i, a_j \} = 0$, $\{a_i^{\dagger},a_j^{\dagger} \} = 0$, $\{a_i ,a_j^{\dagger} \} = \delta_{ij}$, a direct implementation of Eq.~(\ref{eq:Second_quant_Ham}) on a qubit-based quantum processor is not feasible without a mapping from fermionic  to Pauli operators. The fermionic nature of electrons implies that many-electron wavefunctions must be anti-symmetric with respect to particle exchange. This is reflected in the way fermionic creation and annhilation operators act on state vectors: 
\begin{eqnarray}
\hspace{-2cm} a_i^{\dagger} | f_0, \dots, f_{i-1}, f_i, f_{i+1}, \dots, f_n \rangle &= \delta_{f_i,0} \, 
p_i  \,  | f_0, \dots, f_{i-1}, 1, f_{i+1}, \dots, f_n \rangle \label{eq:creat}\\
 \hspace{-2cm}a_i | f_0, \dots, f_{i-1}, f_i, f_{i+1}, \dots, f_n \rangle &=\delta_{f_i,1} \, 
p_i \,  | f_0, \dots, f_{i-1}, 0, f_{i+1}, \dots, f_n \rangle. \label{eq:anhil}
\end{eqnarray}
Here $p_i=(-1)^{\sum_{k=0}^{i-1}f_{k}}$ denotes the parity and $f_i \in \{0,1\}$ the occupation number of the fermionic orbital $i$. The naive replacement of the fermionic operators $a_i^{(\dagger)}$ by Pauli ladder operators $\sigma_i^{\pm} = (\sigma^x\pm i\sigma^y)/2$ does, however, not reproduce Eqs.~(\ref{eq:anhil}) because $\sigma_i^{\pm}$ describe distinguishable {\it particles} with no special symmetries.

A variety of mappings have been developed that guarantee that the fermion statistics are captured on a system of qubits \cite{bravyi_fermionic_2002, tranter_bravyikitaev_2015, bravyi_tapering_2017}. Among those, the Jordan-Wigner mapping \cite{jordan_uber_1928} is particularly intuitive: It is based on a one-to-one mapping of fermionic to qubit occupations, i.e. the occupancy information is stored \emph{locally}. To take into account the parity information $p_i$ in Eqs.~(\ref{eq:anhil}), fermionic operators are translated as
\begin{eqnarray}
 a_i^{\dagger}  \rightarrow \mathds{1}^{\otimes i-1} \otimes \sigma^- \otimes (\sigma^{z})^{\otimes N-i} \\
 a_i  \rightarrow \mathds{1}^{\otimes i-1} \otimes \sigma^+ \otimes (\sigma^z)^{\otimes N-i} \, ,
 \label{eq:JW}
\end{eqnarray}
 where $N$ is the total number of qubits considered. It is obvious that calculating the parity when acting on qubit $i$ requires the knowledge of all state occupations $j<i$, which is accomplished by the $\sigma^z$ terms in Eq.~(\ref{eq:JW}). However, this introduces a non-locality in the mapping and, when inserted into the Hamiltonian in Eq.~(\ref{eq:Second_quant_Ham}), gives rise to long sequences of $\sigma^z$ operators intercalating between $\sigma^{\pm}$ operators of length $k$, known as $k$-local terms. This means that a fermionic wavefunction is spread out over $\mathcal{O}(N)$ qubits, posing fidelity issues in the readout process of the expectation value of the Hamiltonian. 

Recent schemes for tapering off qubits in mapped fermionic Hamiltonians~\cite{moll_optimizing_2016, bravyi_tapering_2017}, based on fermionic symmetries, can partially alleviate the hardware requirements necessary for performing simulations of fermionic systems. These second-quantized tapering schemes exploit symmetries in the mapped qubit Hamiltonian to reduce the simulation space needed to host the mapped fermionic system.

The Jordan-Wigner transformation~\cite{jordan_uber_1928} consists of a local occupancy map and a non-local, $\mathcal{O}(N)$, parity function, whereas the binary-tree transformation encodes both operations on maps that scale $\mathcal{O}(\log(N))$ with the number of qubits~\cite{bravyi_fermionic_2002, tranter_bravyikitaev_2015, bravyi_tapering_2017}, which is a clear advantage compared with the Jordan-Wigner transformation.

\subsection{Coupled cluster trial wavefunctions}
\label{Section:ucc}

Once a mapping of fermions to qubit has been chosen, suitable trial states for the VQE have to be prepared on the quantum processor. At best, these trial states incorporate the structure of the problem Hamiltonian and known properties of the solution state, such as the total number $N$ of fermions. While one could aim to find a gate set that allows one to generate all possible excited Slater determinant configurations, which is known as the full configuration interaction (FCI) approach, the number of states scales factorially with the number of electrons, a clear obstacle for computing larger molecules. One way to improve the efficiency is to use a coupled-cluster approach for creating the trial states, which allows a systematic sampling of all relevant excited Slater determinants up to a given excitation degree. In conventional quantum chemistry, these coupled-cluster expansions are used as a benchmark for all other approaches. 

In the unitary coupled-cluster (UCC) approach \cite{taube_new_2006}, which is a variational version of the commonly used coupled-cluster method \cite{stanton_equation_1993}, the unitary  operator $U(\bm{\theta})$ that is used to generate a trial wavefunction $|\Psi(\bm{\theta})\rangle$ from the reference state $|\Phi\rangle$ is given by
\begin{equation} 
|\Psi(\bm{\theta})\rangle = U(\bm{\theta}) |\Phi\rangle = e^{T(\bm{\theta}) - T^{\dagger}(\bm{\theta})} |\Phi\rangle.
\label{UCC_equation1}
\end{equation}
It is constructed by exponentiation of the cluster operator $T(\bm{\theta})$ defined as 
\begin{equation}
\hspace{-2.4cm}
T(\bm{\theta})  =  \sum_k T_{(k)}(\bm{\theta}) \, , \quad 
T_{(1)}(\bm{\theta})  =  \sum_{i \in {\rm occ} \atop j \in {\rm unocc}} \theta_{(i)}^{(j)} a_j^\dagger a_i  \,  , \quad 
T_{(2)}(\bm{\theta})  =  \sum_{i, j \in {\rm occ} \atop l, k \in {\rm unocc}} \theta_{(i,j)}^{(k, l)} a_l^\dagger a_k^\dagger a_j a_i, ... \quad.  \label{UCC_equation4}
\end{equation}
Here, the coefficients $\bm{\theta}$ describes a vector of parameters that will be optimized using VQE. A common choice for the reference state $|\Phi\rangle$ is the ground-state Slater-determinant made up of the lowest-energy molecular orbitals obtained from the solution of the Hartree-Fock equation. 

The coefficients $\bm{\theta}$ of the cluster operators are not independent and their value decreases with the order of the excitation. Therefore, this expansion is typically truncated at the double (UCCSD) or triple level (UCCSDT) of excitation without significantly reducing the accuracy. In fact, the exponentiation of the cluster operator $T(\bm{\theta})$ introduces higher uncorrelated excitations at each level of truncation, e.~g., for $T(\bm{\theta})=T_{(1)}(\bm{\theta}) + T_{(2)}(\bm{\theta})$
\begin{equation}
e^{T(\bm{\theta})}  =  1 + T_{(1)}(\bm{\theta}) + T_{(2)}(\bm{\theta}) + \frac{T_{(1)}^2(\bm{\theta})}{2!} + T_{(1)}(\bm{\theta}) T_{(2)}(\bm{\theta})  + \frac{T_{(2)}^2(\bm{\theta})}{2!} + ... \, ,
\label{UCC_expT}
\end{equation}
the expansion produced triplet and quadruple excitations in the first few terms of the expansion (fifth and sixth terms, respectively). Despite the compactness of this expansion, the number of coefficients $\bm{\theta}$ increases already in UCCSD with the number of orbitals to the fourth power, which impacts the efficiency of the classical optimization of the trial state $|\Psi(\bm{\theta})\rangle$. In practice, in the case of large molecular systems the limited achievable circuit depth in current quantum devices requires a further truncation of the series in Eq.~(\ref{UCC_expT}). Thus, while the coupled cluster method guarantees in principle an efficient convergence towards the exact ground state, its implementation in state-of-the-art quantum computers requires further studies in terms of how different approximations (truncations) affect the accuracy of the solution.

\subsection{Hardware-efficient trial states suitable for near-term quantum hardware}
\label{sec:heuristic}
 
A much simpler approach is, therefore, the heuristic generation of the trial state with unitary operations that are more suited to the available quantum hardware~\cite{kandala_hardware-efficient_2017}. Independently of the particular problem to be solved, one may choose  trial states that can be efficiently generated in current quantum hardware and at the same time allow the generation of highly entangled states that are close to the target state. 
 
\begin{figure}[tb]
\begin{center}
\includegraphics[width=152mm]{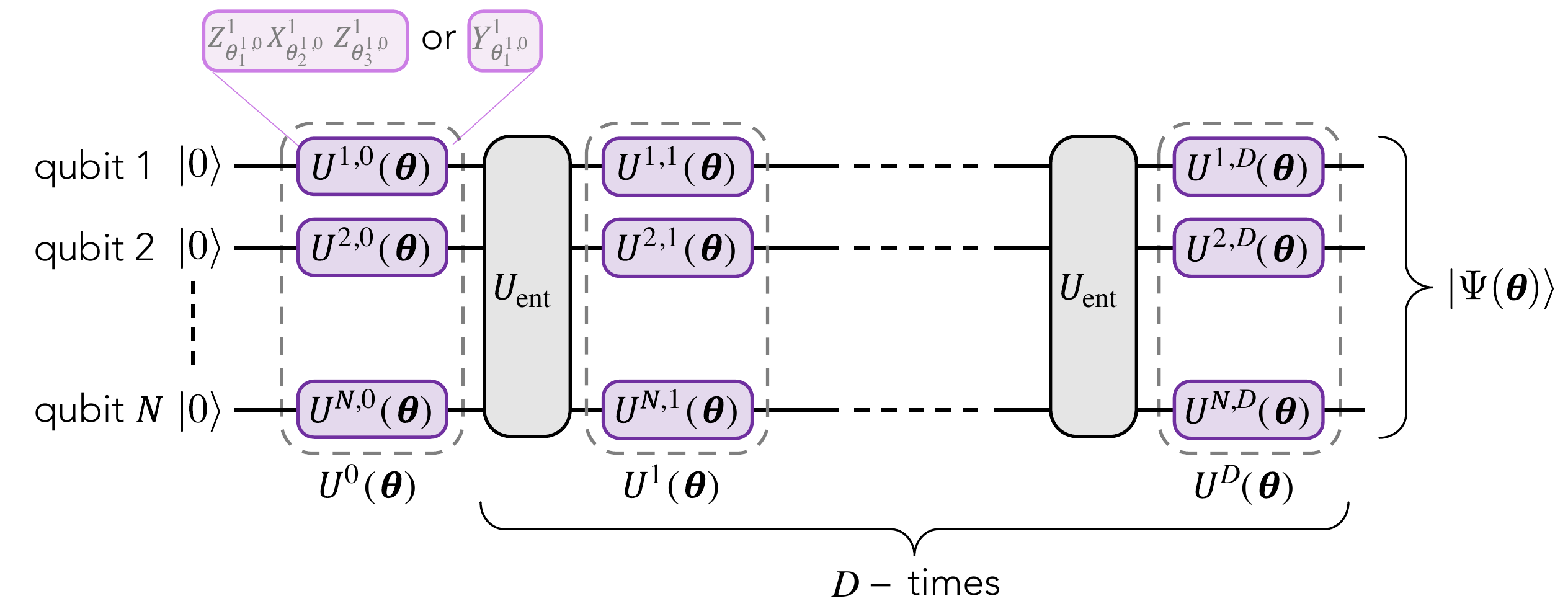}
\end{center}
\caption{\label{fig:heuristiccircuit} Heuristic preparation of trial states for the variational quantum eigensolver based on single-qubit gates $U(\bm{\theta})$ interleaved by entangling operations $U_{\rm{ent}}$ as described in the text.  }
\end{figure}

This approach is showcased in the examples provided in Sections \ref{sec:chemistryexample} and \ref{sec:qaoaexample}. As shown in Fig.~\ref{fig:heuristiccircuit}, the preparation of the heuristic trial states comprises two types of quantum gates, single-qubit Euler rotations $U(\bm{\theta})$ determined by the rotation angles $\bm{\theta}$ and an entangling \emph{drift} operation $U_{\rm{ent}}$ acting on pairs of qubits. 
The $N$-qubit trial states are obtained by applying a sequence of $D$ entanglers $U_{\rm ent}$ alternating with the Euler rotations on the $N$ qubits to the initial ground state $|00\ldots0\rangle$,
\begin{equation}
| \Phi(\bm{\theta}) \rangle = \overbrace{ 
U^{D}(\bm{\theta}) U_{\rm ent}\ldots U^{1}(\bm{\theta})U_{\rm ent}}^{\rm{D-times}} U^{0}(\bm{\theta}) |00\ldots0\rangle
\end{equation}
This gate sequence has a total number of $p = N (3 D + 2)$ independent angles.

To be more specific, the single-qubit operations are decomposed into rotations about the $x-$ and the $z-$axes, $U^{q,i}(\bm{\theta}) = Z^q_{\theta^{q,i}_1}X^q_{\theta^{q,i}_2}Z^q_{\theta^{q,i}_3}$, 
with $X^q(\theta^{q,i}_j) = \exp\left[-i\theta^{q,i}_j\sigma^x_q/2\right]$ (and similarly for $Z^q(\theta^{q,i}_j)$, $Y^(\bm\theta)$) denoting the unitary operation acting on qubit $q$ at the $i$-th position in the gate sequences. The heuristic approach does not rely on the accurate implementation of specific two-qubit gates and can be used with any $U_{\rm ent}$ that generates sufficient entanglement. A natural choice can be the cross-resonance gate~\cite{chow_simple_2011, rigetti_fully_2010} as a two-qubit gate suited for the fixed-frequency superconducting qubit architecture as used, for example, for the IBM Q experience \cite{qx_ibm_quantum_2016}.

\subsection{Small molecules calculated with the variational quantum eigensolver}
\label{sec:chemistryexample}

As an application of the method described above, we present the calculation of the ground-state energy of simple molecules such as the hydrogen molecule: The starting point is the Hamiltonian in second quantization in Eq.~(\ref{eq:Second_quant_Ham}) with the one-body terms, $t_{i j}$, representing the kinetic energy of the electrons and the potential energy that they experience in the presence of the nuclei,
\begin{equation}
\label{eq:t}
 t_{i j}=\int d\boldsymbol x_1\phi_i(\boldsymbol{r}_1) \, \left(-\frac{\boldsymbol\nabla_1^2}{2}+\sum_{n=1}^2 \frac{Z_n}{|\boldsymbol{r}_{1}-\boldsymbol{R}_n |}\right)\phi_j (\boldsymbol{x}_1),
 \end{equation}
 and the Coulomb repulsion terms
\begin{equation}
\label{eq:u}
u_{i j k l}=\int\int d \boldsymbol{r}_1 d \boldsymbol{r}_2 \, \phi_i^*(\boldsymbol{r}_1)\phi_j(\boldsymbol{r}_1)\frac{1}{|\boldsymbol{r}_{1}- \boldsymbol{r}_{2}|}\phi_k^*(\boldsymbol{r}_2)\phi_l(\boldsymbol{r}_2).
\end{equation}
$Z_n$ are the nuclei charges $Z_n$ ($n=1,2$), and each wavefunction $\phi_i(\boldsymbol{x}_1)$ orbital is a $1s$ orbital centered at the one hydrogen atom. We assume that the system is in its spin singlet state. After reduction~\cite{bravyi_tapering_2017} a two-qubit Hamiltonian is obtained 
\begin{equation}
    H_{H_2}= f_0 \, \mathds{1} \otimes \mathds{1}+ f_1 \, \sigma_z \otimes \sigma_z + f_2 \,  \sigma_z \otimes \mathds{1} + f_3 \, \mathds{1} \otimes \sigma_z + f_4 \, \sigma_x \otimes \sigma_x
    \label{eq:Ham_H2_qubits}
\end{equation}
with $f_0=-1.0524$, $f_1=0.01128$, $f_2=0.3979$, $f_3=0.3979$, and $f_4=0.1809$. These coefficients are calculated at the equilibrium distance of 0.74~\AA\ using Eqs.~(\ref{eq:t}) and (\ref{eq:u}).

\begin{figure}[tb]
\begin{center}
\includegraphics[width=100mm]{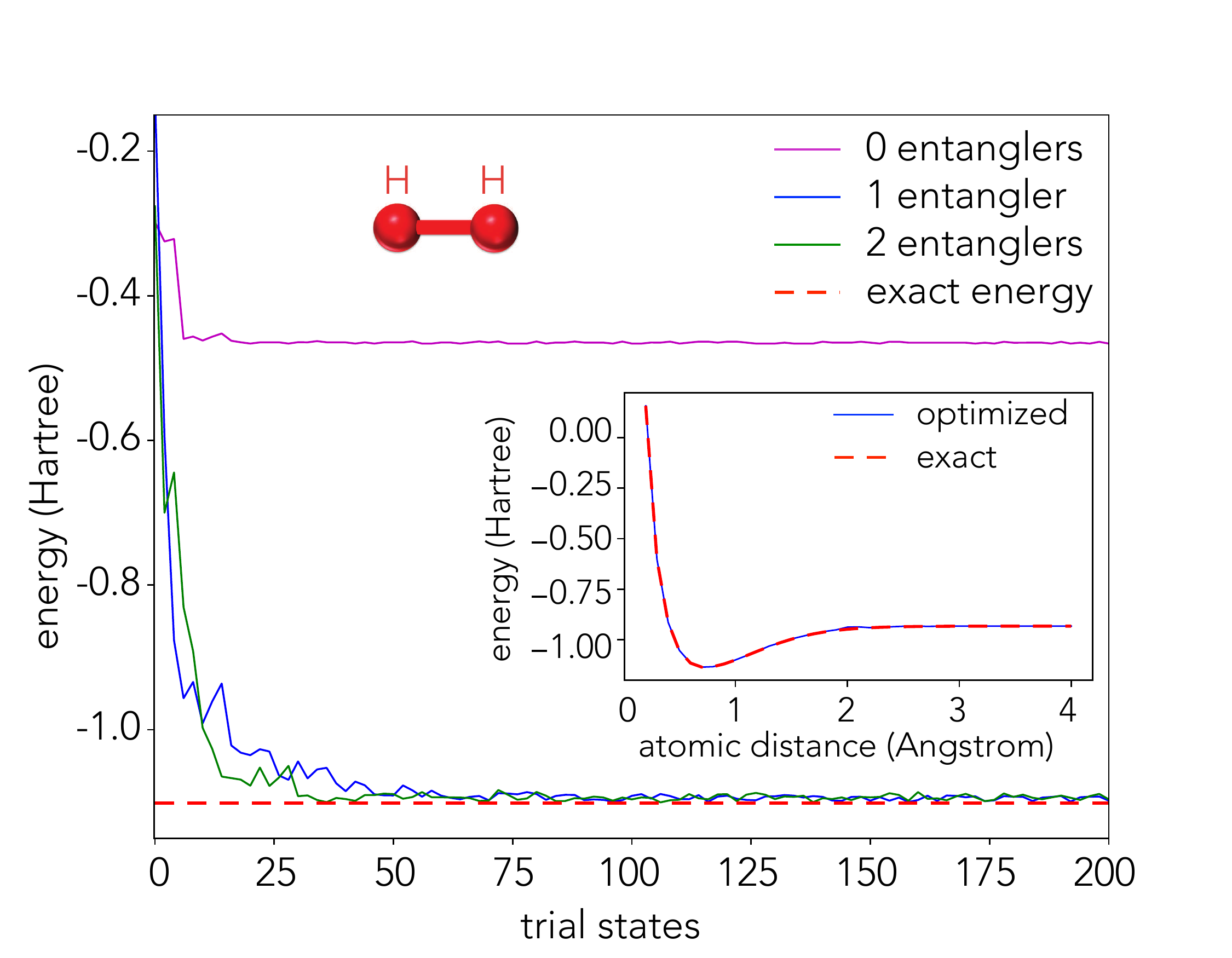}
\end{center}
\caption{\label{fig:qchem} Quantum simulation of the hydrogen molecule on an ideal quantum simulator. At the equilibrium geometry and no entangler block in the circuit, the energy converges to a state with an energy that is about 50\% too high. With two or more entanglers, the exact energy is obtained. The inset shows the entire dissociation profile for a hydrogen molecule calculated with four entangling steps.}
\end{figure}

We evaluate the ground state of the Hamiltonian in (\ref{eq:Ham_H2_qubits}) on an ideal quantum simulator~\cite{qx_ibm_quantum_2016} using a heuristic trial wavefunction approach (Section~\ref{sec:heuristic}) with an increasing number of entangling steps (one, two and four). Here, the single qubit rotations of heuristic trial wavefunctions where implemented as $U^i(\theta) = Y(\theta^i_0) Z(\theta^i_1)$ and the entanglement was introduced via control phase gates~\cite{ibm_qiskit_2017}. Figure~\ref{fig:qchem} shows that a single entangling step is not sufficient to converge towards the correct energy value, whereas two or more entanglers can reproduce the expected results within a few tens of optimization steps in the rotation-angle space $\bm{\theta}$. 

This method can be extended to larger molecules. For lithium hydride (LiH) and beryllium dihydride (BeH$_2$) the second-quantized fermionic Hamiltonian is constructed using a minimal set of atomic orbitals ~\cite{kandala_hardware-efficient_2017} (labelled by the conventional hydrogenic quantum numbers).  In beryllium dihydride the basis is composed of the $1s$, $2s$, $2p_x$ orbitals associated to beryllium, and the $1s$ orbital associated to each hydrogen atom. This results in a total of ten spin orbitals. The two innermost $1s$ spin orbitals of beryllium are assumed to be completely filled. The remaining eight spin-orbitals of beryllium dihydride are reduced to six  by exploiting spin-parity symmetries~\cite{bravyi_tapering_2017}. Similarly, the lithium hydride is mapped onto four qubits. It is demonstrated numerically that in the absence of noise, a number of entangling steps $D = 8$ and $D= 28$ are required to achieve chemical accuracy for lithium hydride and beryllium dihydride, respectively, for the given experimental connectivity. However, the combined effect of decoherence and finite sampling limits the optimal depth for optimizations on current quantum hardware to between zero and two entanglers, which results in deviations of the simulated bond-dissociation energies from the real values. Decreasing the effective error rates or applying error-mitigation schemes as discussed in Section~\ref{sec:error} will  improve the accuracy of the simulations.

\section{Classical optimization with qubits}
\label{sec:qaoa}

The complex Hamiltonians of quantum chemistry problems give quantum computers an inherent advantage over classical hardware. For classical optimization the advantage is not as obvious because many of the relevant problems can be mapped to a relatively simple Ising-spin Hamiltonian. It is diagonal in the computational basis and can be tackled by a range of classical methods. One of the issues with classical solvers is to avoid solutions in local minima of the cost function. In this context simulated annealing~\cite{kirkpatrick_optimization_1983} is an approach that makes use of thermal fluctuations to escape such local minima. Quantum annealing~\cite{kadowaki_quantum_1998} additionally exploits quantum tunneling and can potentially reach a ground state faster especially for problems with very corrugated cost functions~\cite{denchev_whatis_2016, albash_evidence_2017}. The potential for quantum speed-up with this approach is heavily debated in the community; however, because of the tremendous application space even a modest speed-up for a selected number of problems might have a significant impact. Moreover, understanding the detailed evolution of the optimization process and the potential role of entanglement is critical even for improving algorithms that run on classical hardware. This is why the application of the VQE for solving classical optimization problems on gate-based near-term quantum devices is especially interesting. 

To run the variational quantum eigensolver we again consider two different ways to create trial wavefunctions. First, the quantum approximate optimization algorithm (QAOA)~\cite{farhi_quantum_2014} is discussed, which is a polynomial-time algorithm for finding an approximate solution to a classical optimization problem with a desired accuracy. It is related to the quantum adiabatic algorithm~\cite{farhi_quantum_2000}, but has shorter circuit-depth requirements. Second, we give a short example how heuristic trial states can be used to solve a \emph{MaxCut} problem on a real quantum device using the variational quantum eigensolver.

\subsection{Quantum approximate optimization algorithm with short depth}
\label{sec:qaoaexample}

Similarly to the approach described in Section~\ref{sec:heuristic} the trial wavefunction in the QAOA is guided towards the solution by repeated unitary evolution according to two Hamiltonians. The first one is the Hamiltonian $H_C$, which encodes the classical cost function $C(\mathbf x)$ of a binary constrained optimization problem. The second one is a mixing Hamiltonian $H_M$, which helps guide the optimization in Hilbert space towards the ground state of $H_C$.  The number of times that both Hamiltonians are applied in the optimization process defines the level $D$ of the circuit and determines the complexity of the algorithm. 

Without loss of generality, it is assumed that an optimal solution $\mathbf x$ \emph{minimizes} the cost function $C(\mathbf x)$ which is a polynomial in the binary components $x_i \in \{ 0,1\}$ of the variable $\mathbf x$. Encoding of the cost function $C(\mathbf x)$ into a Hamiltonian $H_C$ requires shifting the binary variables $x_i\to(1-z_i )/2$ with $z_i \in \{ -1,1\}$ and then substituting $z_i\to\sigma_i^z$ to obtain an Ising-type Hamiltonian. We chose the same notation as in Eq.~(\ref{eq:Hqpauli}) but consider only diagonal terms $\sigma_i^j \in \{\mathds{1},\sigma_i^z\}$ which gives
\begin{equation}
H_C = \sum_\alpha h_\alpha P_\alpha = \sum\limits_{\alpha} h_\alpha \bigotimes_{i_\alpha} \sigma_{i_\alpha}^z \, . 
\label{eq:QAOA_Hamil}
\end{equation}
Here the index $i_\alpha$ runs over all $\sigma_{i_\alpha}^z$ in $P_\alpha$, which constitutes a $k$-local term (many-body interaction term among $k\leq N$ qubits), matching the polynomial terms in the cost function $C$ with corresponding real coefficients $h_\alpha$. The second Hamiltonian $H_M$ is just a global transverse field, i.e.\ $H_M = -\sum\limits_i \sigma^x_i$. To find the ground state of the problem Hamiltonian $H_C$, one proceeds by applying the evolution operator 
\begin{equation}
U(\bm{\beta}, \bm\gamma)=\prod\limits_{l=1}^{D}e^{-i\beta_l H_M}e^{-i\gamma_l H_C}
\end{equation}
to a starting state $|\psi_0\rangle$ that can easily be generated on the quantum computer, e.~g.\ a uniform superposition state. Using the VQE, the parameters of the final state $| \bm{\beta}, \bm\gamma \rangle = U(\bm{\beta}, \bm\gamma)|\psi_0\rangle$ are then adjusted such as to minimize the expectation value  $\langle \bm{\beta}, \bm\gamma|H_C | \bm{\beta}, \bm\gamma \rangle$. Measurement of the final state $| \bm{\beta}, \bm\gamma \rangle$ directly yields the solution of the classical optimization problem with a probability that approaches unity as $D$ increases. However, with increasing $D$ the circuit depth required will reach the decoherence limits of available quantum hardware, and the fidelity of the result will again decrease. Also, the number of classical parameters that need to be optimized for large $D$ will result in a slower convergence. Instead of using the VQE choosing a fine interpolation $(\beta_l,\gamma_l)=(1-l/D,l/D)$ with $l= 0,...,D$ would be equivalent to first order with a trotterized version of the adiabatic quantum algorithm~\cite{lloyd_universal_1996, farhi_quantum_2000}. By letting the VQE select optimal parameters $(\gamma_l,\beta_l)$, a more direct path to the target state becomes possible and the algorithm can reach the ground state with high accuracy even for relatively small values of $D$. The QAOA has been generalized and successfully applied to MaxCut with analytical and numerical studies~\cite{farhi_quantum_2017}.

\subsection{Variational quantum eigensolver applied to the MaxCut problem}
\label{Section:maxcut}

To give an example of a classical optimization problem, we discuss an instance of the maximum-cut (MaxCut) problem with five qubits. Instead of generating trial states with the QAOA, we again use the hardware-efficient approach explained in Section~\ref{sec:heuristic} to run the algorithm on a real quantum device. The MaxCut problem is an NP-complete binary optimization problem, with applications in clustering, network science, and statistical physics. It aims at grouping the nodes of a graph into two subgroups by cutting across the links between them. The cut is to be made in such a way that the added weights of the links (edges) that were cut are maximized.

\begin{figure}[tb]
\begin{center}
\includegraphics[width=122mm]{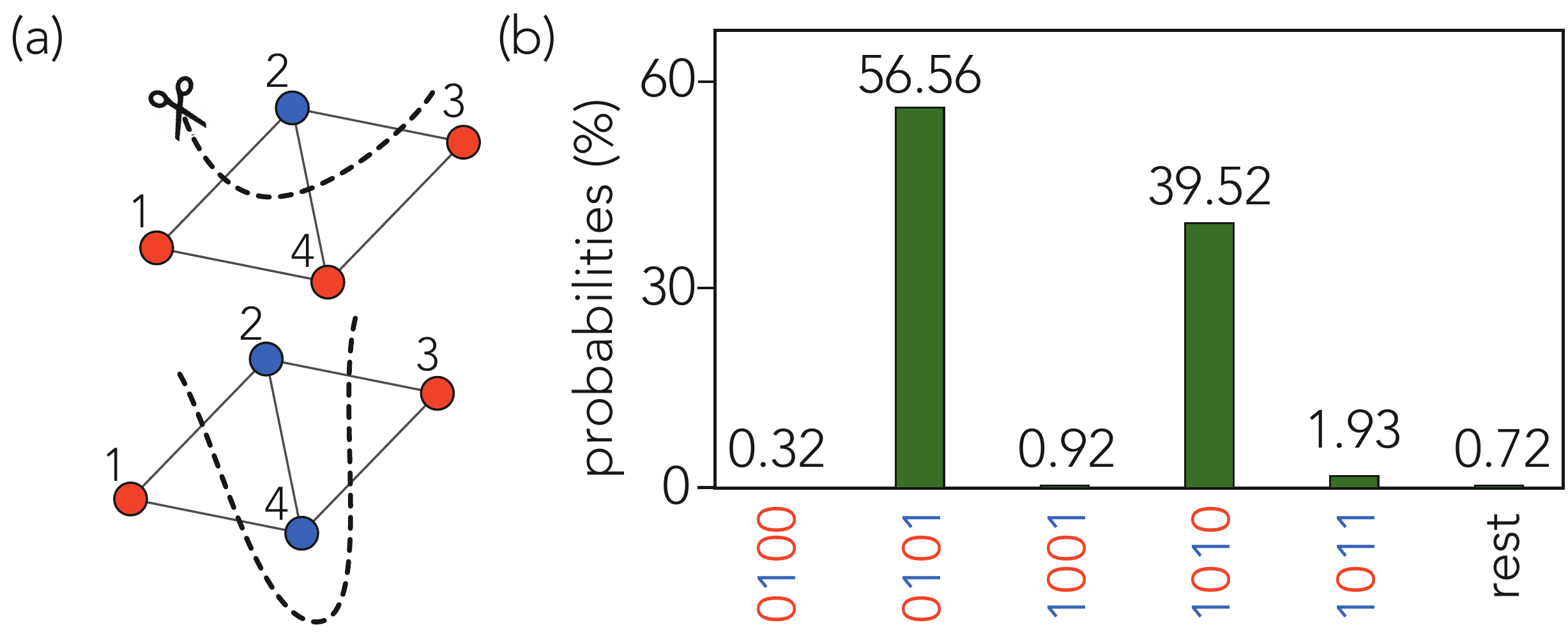}
\end{center}
\caption{\label{fig:maxcut}(a) Two different cuts for a MaxCut instance with four vertices. The lower cut has a larger cumulated weight and represents the partitioning that solves the problem. (b) Solution of the problem using the VQE with heuristic trial-states and a depth $D = 3$ circuit.}
\end{figure}

The formal definition of this problem is the following: Consider an $n$-node non-directed graph with edge weights $w_{ij} > 0$, $w_{ij} = w_{ji}$, where $(i, j)$ enumerate the nodes linked by the corresponding edge~\cite{lucas_ising_2014}. The profit function to be maximized is therefore the sum of edge weights connecting points in the two different subsets. By assigning a subset label $x_i=0$ or $x_i=1$ to each node $i$, one tries to maximize
\begin{equation}
C(\textbf{x}) = \sum_{i,j} w_{ij} x_i (1 - x_j) \, .
\end{equation}
We can then use the mapping described in Section~\ref{sec:qaoaexample} to obtain the Ising Hamiltonian
\begin{eqnarray}
H_I & = & \sum_{i<j} \frac{w_{ij}}{2} (1-\sigma^z_i)(1+\sigma^z_j) \\
& = &  -\frac{1}{2}\sum_{i<j} w_{ij} \sigma^z_i \sigma^z_j+\mathrm{const} \, ,
\end{eqnarray}
In other words, the weighted MaxCut problem is equivalent to finding the ground state of the Ising Hamiltonian
\begin{equation}
H_C = \sum_{i<j} w_{ij} \sigma^z_i \sigma^z_j \, .
\end{equation}
For exploring the solution space of $H_C$ we use the approach from Section~\ref{sec:heuristic} to define a hardware-efficient heuristic trial wave function 
\begin{equation}
|\psi(\theta)\rangle = [U(\boldsymbol\theta) U_\mathrm{ent}]^D |\psi_0 \rangle \, ,
\end{equation}
where $U_\mathrm{ent}$ is a collection of fully entangling gates that are diagonal and the number of entanglers $D$ defines the level of the quantum circuit. The single-qubit rotations are chosen to be $U(\theta) = \prod\limits_{i=1}^N Y(\theta_{i})$, where $N$ is the number of qubits. For a classical problem this choice allows a search over the space of quantum states with only real coefficients, while still exploiting entanglement to potentially converge faster to the solution. Evaluation of the energy expectation value for a specific trial wavefunction is especially simple in this case as it is sufficient to measure all four qubits and extract the pairwise $\sigma^z_i \sigma^z_j$ correlators. Figure~\ref{fig:maxcut}(a) shows two different cuts through a problem instance with four nodes (qubits). The lower of the two solves the problem if all non-zero weights in $w_{ij}$ are assumed to be equal. 
When we implement this on an ideal quantum simulator~\cite{qx_ibm_quantum_2016, ibm_qiskit_2017} and use the VQE to optimize the parameters of the trial state in 100 trial steps, we get the state probabilities shown in Fig.~\ref{fig:maxcut}(b). For this simple simulation, the solution is found with a probability that is higher than $95\%$.

\section{Classical robust optimizers for measured expectation values}
\label{sec:opt}

The optimization cycle of the VQE (see Section \ref{sec:vir}) involves evaluation of the cost function on a real quantum device, e.~g., a superconducting quantum processor, and adjustment of the variational parameters using classical optimization algorithms (see Section~\ref{sec:vir}). In the latter, several important aspects need to be considered for successful application of the VQE.

First, the optimization could get stuck in a local minimum  that would correspond to an excited state of the system. Using a suitable optimization routine can prevent finding such false minima. Gradient-descent methods may be combined with simulated annealing steps or strategies that involve starting from multiple initial points. In this context, in~\cite{wecker_progress_2015} a greedy search with multiple starting points is alternated with a Powell search, showing good performances on Hubbard lattices of up to twelve sites.

Second, because of the limited number of samples of the Hamiltonian terms on the quantum computer one only has access to a noisy energy (cost) value. The error in the energy estimation goes as $\mathcal{O}(1/\sqrt{s})$, with $s$ the number of samples taken.  Grouping  Pauli operators into commuting sets \cite{mcclean_theory_2016, kandala_hardware-efficient_2017} that can be measured with the same state preparation and post-rotations reduces the number of separate measurements and enables more averages and better sampling statistics. Still, the choice of the optimizer must take into account that the cost function is affected by stochastic fluctuations. In fact, while unitary coupled-cluster methods and other analytical variational circuits in principle support the use of gradient-based methods that increase the efficiency of the optimization~\cite{romero_strategies_2017}, an imperfect knowledge of the unitary gates implemented in a given quantum device and statistical noise render gradient-based approaches less useful. Derivative-free methods, such as Nelder-Mead and the TOMLAB method, have been tested for optimization of the hydrogen molecule, resulting in a superior performance of the latter method in the presence of stochastic noise~\cite{mcclean_theory_2016}. 

Third, time overheads due to repeated sampling and the number of function evaluations to update the variational parameters will affect the performance of the optimization. In this spirit, the use of a simultaneous perturbation stochastic approximation (SPSA)~\cite{spall_multivariate_1992}, used in~\cite{kandala_hardware-efficient_2017} for molecular structure problems, provides both a constant overhead in terms of the number of variational parameters and robustness with respect to stochastic fluctuations. Extensions of the SPSA method that include approximations to the Hessian matrix can be explored to improve the speed of the optimization in the final steps, where estimating second derivatives helps achieve faster convergence~\cite{spall_adaptive_2000}. In contrast, additional savings in time overhead in SPSA optimizations that rely on just one evaluation of the cost function per update step~\cite{spall_one-measurement_1997} could further improve the performance in large-scale quantum problems where sampling is particularly difficult. While simultaneous perturbation methods can be very useful in the optimization of fermionic problems, for classical problems, such as instances of MaxCut, the ease of evaluating the cost function may favor standard gradient-descent or derivative-free routines. 

Another critical aspect is the improvement of the classical control hardware for running the VQE on a quantum device: measurement of the cost function with sufficient accuracy requires repeated sampling of the output state and thereby also repeated cycles of qubit initialization, application of the quantum gates and qubit measurement. The speed of the execution of the optimization can be improved on the hardware side by using integrated active reset techniques. In the case of superconducting qubits this is true for both qubits and resonators~\cite{mcclure_rapid_2016, bultink_active_2016}. Moreover, the costly time overhead in synthesizing and loading control pulses onto the quantum processor for trial-state preparation can be reduced by short-latency field-programmable gate-array-based control and measurement architectures such that time overheads are solely related to the execution of the quantum gates and the readout of the qubits.

\section{Prospects of fighting decoherence without full error correction}
\label{sec:error}

The hardest challenge for practical near-term quantum devices is their sensitivity to noise. Any computation that has the potential to leverage quantum effects and to provide a quantum speed-up over classical algorithms needs sufficiently coherent qubits. It was realized early on~\cite{unruh_maintaining_1995} that the coupling to the environment sets both a time and size limit for a quantum computation. Hence, the strength of this coupling determines how large a computation can be performed. This constant limit has to be contrasted to the improvements that are gained from the asymptotic scaling advantages of quantum algorithms. This limitation was, at least in theory, remedied with the advent of quantum error correction~\cite{shor_scheme_1995, steane_error_1996, calderbank_good_1996}. However, in spite of rapid experimental progress, the resource requirements for fully fault-tolerant operations with current codes~\cite{fowler_surface_2012} seem prohibitively large~\cite{jones_layered_2012, devitt_requirements_2013}. In turn, hopes were raised that non-error-corrected devices will soon become available that reach a regime of reasonably long coherence times and give rise to dynamics too complex to be simulated on a classical computer~\cite{boixo_characterizing_2016, farhi_quantum_2016}. In light of these developments, the question arises which computational tasks can be accomplished with quantum devices that have only limited or no error correction. Depending on the form of the actual physical noise, it is expected that the production of entropy in any quantum circuit that is subject to noise will set a limit to this approach~\cite{aharonov_limitations_1996}, and error correction is indispensable for any advanced form of quantum information processing. However, the full computational power of even short-depth circuits is not yet fully understood, and based on complexity-theoretic grounds, it can be argued, that even finite-depth circuits lie beyond the computational power of a classical computer~\cite{terhal_adaptive_2002, farhi_quantum_2016}.

Recent experiments in which the quantum simulation of small molecules was performed~\cite{kandala_hardware-efficient_2017} showed that even for very short-depth circuits the effects of decoherence become apparent. For the simulation to be of value, the effect of this error needs to be mitigated, and several proposals have been made to deal with the effects of decoherence in short-depth quantum computation~\cite{mcclean_hybrid_2017, li_efficient_2017, temme_error_2016, schwenk_reconstructing_2017}.

For a large fraction of applications, the computational task can be abstracted to estimate the expectation value of some observable after the application of a short-depth quantum circuit. This estimation must be accurate enough to achieve a simulation precision that outperforms approximate classical simulation tasks. Techniques to mitigate the error in the estimation of expectation values were introduced in \cite{temme_error_2016}. It is shown that the estimate can be improved in the presence of noise with only a modest time overhead. This approach requires no additional hardware resources such as fresh ancilla or code qubits.

In this scheme, the estimation of an expectation value is improved by an \textit{extrapolation to the limit of zero noise} as originally proposed by Richardson~\cite{richardson_deferred_1927}. The method requires no \textit{a priori} knowledge about the noise source, except that the noise is weak and time-independent. To understand this approach it is useful to choose a more physically motivated description of the computation rather than the gate-based quantum circuits. It is more convenient to consider a time-dependent Hamiltonian dynamics $H(t)= \sum_{\alpha}{J_{\alpha}(t) P_{\alpha}}$ that implements the circuit, where $J_{\alpha}(t)$ are coupling coefficients and $P_{\alpha}$ are $N$-qubit Pauli operators. In this model the coherent evolution is subject to a noise contribution ${\cal L}$ that is effectively constant in time and acts on a time scale much larger than the time-dependent Hamiltonian implementing the quantum circuit. The time evolution up to some time $T$ of the open system with initial state $\rho_0$ can by described by a Lindblad master equation
\begin{equation}
	\frac{\partial}{\partial t} \rho(t) = -i[H(t),\rho(t)] + \lambda {\cal L}(\rho(t)) \, .
\end{equation}
The expectation value $E(\lambda)$ of some observable $A$ is then obtained from the final state $\rho_\lambda(T)$ and can be written as a power series of the noise rate $\lambda$
\begin{equation}
E(\lambda)= E^{*}(0)+ \sum_{i=1}^{n}{a_i \lambda^{i}} + O(\lambda^{n+1})
\end{equation}
where $E^{*}(0)$ corresponds to the noise-free expectation value.

Richardson proposed a so-called {\em deferred approach to the limit} to estimate an expectation value such as $E^{*}(0)$ with high accuracy~\cite{richardson_deferred_1927, sidi_practical_2003}. For this purpose, the expectation value $E(\lambda_j)$ is measured for different noise rates $\lambda_j=c_j \lambda$, where $c_j$ is a rescaling factor and $\lambda$ the actual noise rate in the experiment. The noise-free expectation value can then be estimated by~\cite{temme_error_2016}
\begin{equation}
E^{*}(0) =  \sum_{j=0}^{n}{\gamma_j E(\lambda_j)} + O(\lambda^{n+1})
\end{equation}
where $\sum_{j=0}^{n}{\gamma_j} = 1$ and $\sum_{j=0}^{n}{\gamma_j c_j^k} = 0$ for $k=1...n$. In this way the largest terms in the error up to $O(\lambda^n)$ are cancelled, thus leading to an estimation of the noise-free expectation value with very high accuracy. In practice however, the noise rate $\lambda$ is fixed. To still obtain an experimental estimate of the expectation values $E(\lambda_j)$, the following trick can be applied: the quantum circuit $H(t)$ can be run for a time $c_j T$ and with a reduced coupling $J_{\alpha}/c_j$. As the noise ${\cal L}$ is assumed to be constant in time, it can be shown that the state resulting from a rescaled dynamics is identical to the state obtained from the dynamics with an effectively rescaled noise parameter. Depending on the nature of the noise, relative errors for the noise-free expectation value range from $10^{-6}$ to $10^{-11}$~\cite{temme_error_2016}.

\section{Conclusion}
\label{sec:conclusion}

Current and near-term quantum processors will most likely be limited to a few hundred, maybe a thousand qubits, and operate without quantum error correction. If the qubits and their control were ideal, the computational power of quantum devices with a couple hundred qubits would already dwarf that of any classical computer and could show \emph{quantum advantage}. However, errors in the quantum operations reduce their computational power.

In this paper it is argued that a proper metric, such as the \emph{quantum volume}, should be used to assess the computing power of a quantum processor and to compare different prototypes on a fair basis. With this metric, it becomes clear that not only the qubit number has to be increased, but also and even more importantly, the effective error rate needs to be significantly reduced before practical applications come within reach. Simple estimates show that to run a algorithm with depth hundred on a hundred-qubit device requires an effective error rate of 0.01~\%. This number is not completely unrealistic, but shows the necessity to construct algorithms with short depth. Moreover, error-mitigation schemes using no or only a small number of extra ancilla qubits will be important to compensate systematic deviations in the computed result.

Besides enlarging the quantum volume and reducing the effect of errors, it is essential to find suitable methods and algorithms to use quantum effects efficiently. We have discussed that a promising way forward is to consider hybrid quantum-classical architectures in which the quantum processor is used to generate trial quantum states that could not be stored in conventional memory. The variational quantum eigensolver method can be use to solve any type of problem that can be cast into a physical Hamiltonian. Constrained binary optimization problems can be described by an Ising-type Hamiltonian, whereas problems from the field of quantum chemistry or material science map into a more general spin Hamiltonian with more than longitudinal interactions among the spins.

\begin{figure}
\begin{center}
\includegraphics[width=140mm]{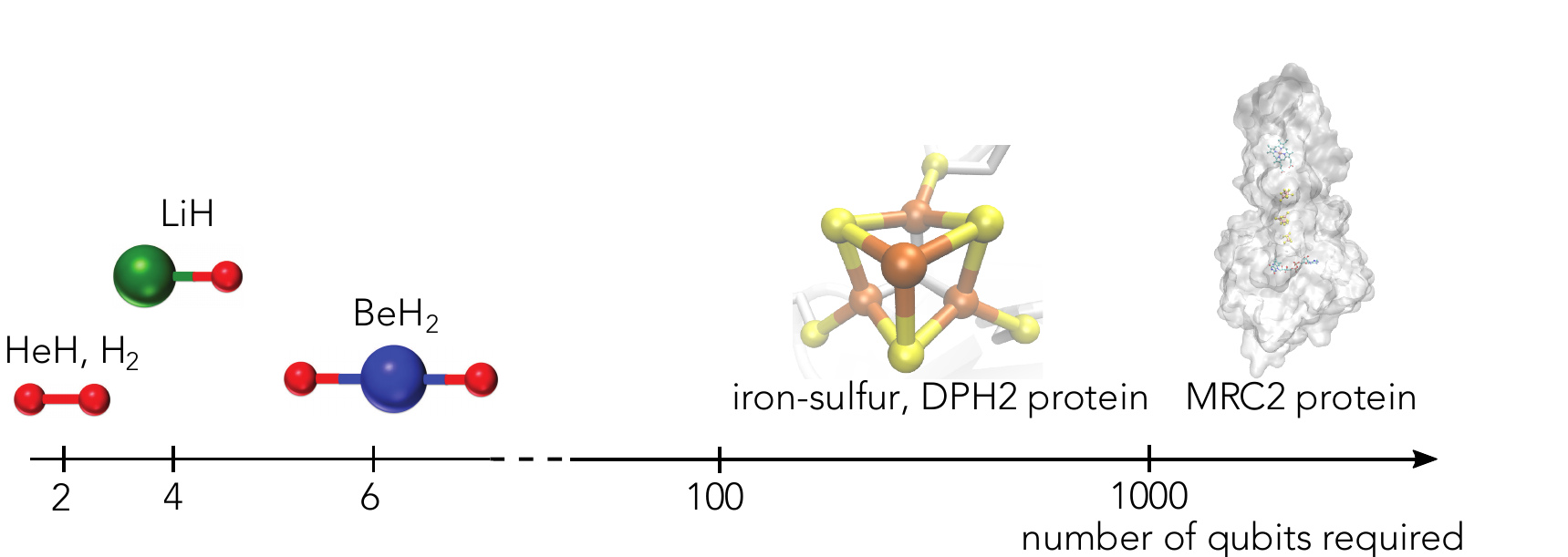}
\end{center}
\caption{Qubit resources needed for quantum chemistry. Qubit numbers up to ten are based on existing experiments, whereas the resources for larger molecules are estimates. From left to right: hydrogen molecule, lithium hydride, beryllium hydride, iron sulphor (Fe-S) cluster in DPH2 complex of Pyrococcus Horikoshii (PDB entry code 3LZD), and Fe-S clusters sequence in cytochrome B560 subunit of mitochondria (PDB entry code 3SFD).}
\label{fig:QCqubits}
\end{figure}

For Ising-type Hamiltonian problems, it is not clear how much quantum speed-up can be expected, because many fast classical algorithms have already been developed~\cite{farhi_quantum_2014}. In contrast, the Hamiltonian for chemistry and materials-related problems contains so-called non-stoquastic terms, which makes it difficult to solve these problems exactly on a classical computer. It is, therefore, believed that using a quantum processor will lead to exponential speed-up. The current state of the art encompasses proof of concept simulations of small molecules: In the context of superconducting qubits the hydrogen molecule has been simulated with two qubits~\cite{omalley_scalable_2016, colless_robust_2017, kandala_hardware-efficient_2017} and larger molecules such as lithium hydride and beryllium dihydride have been simulated with seven qubits~\cite{kandala_hardware-efficient_2017}. As the size of the systems under study grows in electron number so does the required number of qubits, for example, the simulation of the electronic structure of small organic molecules such as benzene and ethane~\cite{kassal_simulating_2011} already requires tens to hundreds of qubits. In the case of strongly correlated electrons, even the simplest systems made of a few atoms, like for instance the chromium dimer~\cite{booth_linear_2014}, quickly become intractable for classical computers when accurate numerical solutions are required. To address strongly correlated problems of practical relevance such as the nitrogen fixation catalytic center in bacteria~\cite{reiher_elucidating_2017} or the iron-sulphur clusters in the respiratory chain protein complexes~\cite{zhang_diphthamide_2010, zhou_thiabendazole_2011} (see Figure~\ref{fig:QCqubits}) quantum processors with a significantly increased quantum volume are needed.
To achieve this, the capabilities of next-generation quantum processors have to improve along several directions:
\begin{enumerate}
\item \label{item:errors} Improvement of coherence and qubit control, as well as development of error-mitigation schemes.
\item \label{item:trialstate} Hardware-efficient and problem-specific trial state preparation when using variational quantum eigensolver.
\item \label{item:fermion} Efficient circuit optimization by code optimizers and improved mappings from fermions to qubits. 
\item \label{item:optimization} Classical parameter optimization methods suited for variational quantum eigensolver.
\end{enumerate}

As for (\ref{item:errors}), current best error rates of $\sim10^{-4}$ for single and $\sim10^{-3}$ for two-qubit gate fidelities in the case of superconducting qubit architectures do not provide sufficient accuracy for more complex quantum calculations. The coherence time of qubits has to be improved, e.~g., by improving fabrication techniques or chip designs. The control pulses for qubits and their interaction have to be optimized to avoid systematic gate errors. Any remaining errors have to be compensated by error-mitigation strategies.

As for (\ref{item:trialstate}), trial states that require only a  variation of a few parameters to prepare the targeted solution state are required. It is an open question how to construct suitable trial states for a general problem set. One may speculate that some combination of heuristic and problem-specific approaches is best suited for the variational quantum eigensolver, e.~g., hardware-efficient trial wavefunctions which obey certain physical constraints, for example, to conserve the particle number in the quantum chemistry context. Moreover, enlarging the set of available gates, e.~g.\ by exploring coupling primitives that allow different types of interactions between two or more qubits to be realized~\cite{mckay_universal_2016, roth_analysis_2017} is considered to create problem-specific trial states and render the VQE efficient.

As for (\ref{item:fermion}), different fermions-to-qubits maps have been proposed which do not require the creation of entanglement over the entire qubit space. Among the different variants of the Jordan-Wigner and binary-tree methods, one can envisage approaches that perform better in the presence of system-specific noise. Moreover, it may be possible to identify new maps into qubits, which are especially suited for variational quantum eigensolvers and that can exploit, for instance, the use of additional ancilla qubits to further reduce the number and the complexity of the gates. Of particular interest is also the possibility to optimize quantum circuits using {\em post-processing} tools at compilation~\cite{reiher_elucidating_2017}. The use of high-level languages for the generation and the manipulation of quantum circuits will indeed offer the possibility to rationalize the qubits resources, thus reducing the circuit depth and therefore the time to solution.

As for (\ref{item:optimization}), specialized classical optimizers that can deal with large stochastic fluctuations resulting from queries to the quantum processor in the VQE are required. The possibility that optimization routines get trapped in false local minima or the effect of high noise render the robustness of optimizers of critical importance for near-term applications. Even the use of quantum-enhanced optimization schemes may be envisaged.

In conclusion, several promising approaches to make use of near-term devices with hundreds of qubits and limited coherence times have been developed. Overcoming the remaining challenges will allow us to solve tangible problems, most likely in quantum chemistry, material science or classical optimization.



\section*{References}

\bibliographystyle{iopart-num}

\bibliography{1000qubit}

\end{document}